# From a superconductor NdNiO$_2$ to a Mott multiferroic BiNiO$_2$

Hu Zhang*, RuiFeng Zhang, Lulu Zhao, Chendong Jin, Ruqian Lian, Peng-Lai Gong, RuiNing Wang, JiangLong Wang, and Xing-Qiang Shi

Key Laboratory of Optic-Electronic Information and Materials of Hebei Province, Research Center for Computational Physics, College of Physics Science and Technology, Hebei University, Baoding 071002, P. R. China

* E-mails: zhanghu@hbu.edu.cn

Motivated by the recently discovered superconductivity in Sr-doped nickelate oxides NdNiO$_2$, we predict a material BiNiO$_2$ that provides an opportunity to study the intertwined ferroelectricity, metallicity, and magnetism in a crystal with very simple atomic structures. There is a ferroelectric structural transition from the nonpolar phase with the *P*4/*mmm* space group to the polar phase with the *P*4*mm* space group, which is driven by the lone pair on Bi. Calculations based on the Heyd-Scuseria-Ernzerhof hybrid density functional reveal that both the nonmagnetic and ferromagnetic states are metallic for nonpolar and polar phases, while the lowest energy ground-state for polar BiNiO$_2$ is a Hubbard Mott insulator with the G-type antiferromagnetic spin configurations. As a ferroelectric material with an electric polarization of 0.49 C/m$^2$, it may be possible to control the magnetic order in BiNiO$_2$ by an applied electric field. The replacement of Nd by Bi serves as a connecting link between a high-temperature superconductor and a Mott multiferroic. Our work supports a route towards strongly correlated ferroelectrics.

## I. INTRODUCTION

There are two important types of coupled physical properties in ferroelectrics. One is the interplay between the ferroelectricity and magnetism [1]. Such materials are usually called multiferroics. The coexistence of ferroelectric and magnetic order in multiferroics provides a route for the control of magnetism by applications of an electric field. This magnetoelectric coupling makes these materials be useful for technological applications including sensors, spintronics, multifunctional memory, and so on. BiFeO$_3$ is one of the typical multiferroics having magnetic and ferroelectric order at room temperature [1]. The other of the two types of coupled physical properties in ferroelectrics is the interplay between the ferroelectricity and metallicity. At first sight it seems that metallic materials cannot exhibit ferroelectricity because conduction electrons will screen internal electric fields [2]. In 1965, Anderson and Blount predicted the possibility of a "ferroelectric metal" [3]. In other words, there was a ferroelectric-like structural transition in the metallic material. In 2013, a ferroelectric-like structural transition in the metallic LiOsO$_3$ was discovered experimentally [4]. Such materials are also called polar metals due to its ferroelectric-like polar crystal structures. Other works on polar metals have also been published [5-7]. Elemental polar metals (group V elements crystallize in the *P*6$_3$*mc* space group) were also predicted [8]. From a fundamental standpoint, the coupling of disparate physical properties including ferroelectricity, metallicity, and magnetism in a single phase is very interesting.

In this work, we theoretically predict a material BiNiO$_2$ which allows us to study the intertwined ferroelectricity, metallicity, and magnetism, as shown in Fig. 1, in a crystal with very simple atomic structures. This prediction is partly motivated by the recently discovered superconductivity in Sr-doped nickelate oxides NdNiO$_2$ having the *P*4/*mmm* (No. 123) space group [9]. In experiments, the infinite layered phase NdNiO$_2$ was obtained from the reduced form of perovskite nickelates NdNiO$_3$. Similar to NdNiO$_2$, we may expect that BiNiO$_2$ can be synthesized experimentally from the reduced form of perovskite nickelates BiNiO$_3$. Our results indicate that there are one unstable polar-mode (imaginary frequencies of phonons) in BiNiO$_2$ with the *P*4/*mmm* symmetry due to the lone pair on Bi. According to the soft-mode theory of ferroelectrics, this unstable phonon drives a ferroelectric structural transition, in which nonpolar BiNiO$_2$ transforms into a polar state with the *P*4*mm* (No. 99) space group without the inversion symmetry. Similar to NdNiO$_2$, BiNiO$_2$ has the strongly correlated electronic structures. We find that both the nonmagnetic and ferromagnetic states of the nonpolar and polar $BiNiO_2$ are all metallic. The G-type antiferromagnetic (AFM between $NiO_2$ planes along the *c* direction and AFM within the $NiO_2$ plane) state has the lowest ground-state energy for polar $BiNiO_2$. In fact, the polar phase of $BiNiO_2$ is a Mott insulator. Therefore, the ferroelectric state of $BiNiO_2$ provides a good platform to study various coupled physical properties.

## II. METHODOLOGY

We have performed first-principles calculations based on density functional theory (DFT) [10] with the local density approximation (LDA) and Perdew–Burke–Ernzerhoff (PBE) functional in generalized gradient approximation (GGA) [11] in the Vienna Ab Initio Simulation Package (VASP) [12-14]. The Heyd-Scuseria-Ernzerhof (HSE) hybrid functional [15] was used to obtain the strongly correlated electronic structures. We used a 12×12×12 Monkhorst-Pack grid [16] in the LDA and PBE calculations with an energy cutoff of 500 eV. For HSE calculations, a 6×6×6 Monkhorst-Pack grid was used. The crystal structures were relaxed until the Hellmann-Feynman forces are less than 1 meV/Å. The Berry phase method [17] was used to calculate the electric polarization. The phonon spectra were calculated with phonopy [18]. The tight-binding models were constructed with the maximally localized Wannier functions using Wannier90 package [19,20]. The $k · p$ method was used to obtain the effective Hamiltonian describing the quadratic band-crossing point (QBCP).

## III. RESULTS AND DISCUSSION

### A. The ferroelectric structural transition

According to experimental results, $NdNiO_2$ has the *P*4/*mmm* space group with the lattice constants *a* = 3.9208 Å, *c* = 3.281 Å. Nd, Ni, O1, and O2 atoms site at the 1*d* (1/2, 1/2, 1/2), 1*a* (0, 0, 0), and 2*f* (0, 1/2, 0) (1/2, 0, 0) Wyckoff positions respectively [21]. These crystal structure parameters are used as the initial input for the nonpolar $BiNiO_2$ and then we relax them fully. For the LDA calculations we obtain *a* = 3.835 Å, $c$ = 3.324 Å. While PBE gives $a$ = 3.931 Å, $c$ = 3.395 Å. It is well known that LDA and PBE usually underestimate and overestimate the lattice constants slightly, respectively, compared to the experimental values. The calculated structural parameters for nonpolar $BiNiO_2$ are collected in Table I. Our calculation results indicate that $BiNiO_2$ and $NdNiO_2$ have similar lattice constants. We find that the LDA and PBE crystal structure parameters give nearly identical electronic structures. Hence, we will use the crystal structure parameters obtained with LDA calculations in the following. The atomic structures of nonpolar $BiNiO_2$ are shown in Fig. 2(a), in which four O atoms surround the Ni in a planar square environment.

According to the soft-mode theory of ferroelectrics, there exist unstable polar phonons in the nonpolar high-symmetry reference structure in first-principles calculations [1]. From the factor group analysis, the nonpolar $BiNiO_2$ with *P*4/*mmm* symmetry has 3 acoustic modes and 9 optical modes at the Brillouin zone center. Optical modes can be expressed as

$$3E_u + B_{2u} + 2A_{2u}, \tag{1}$$

where the three two-fold degeneracy $E_u$ modes and two $A_{2u}$ modes are infrared active. The calculated frequencies of modes at the Brillouin zone center are given in Table II. We find one unstable polar-mode with an imaginary frequency of 215 cm$^{-1}$ (denoted as a negative value). This is one of the $A_{2u}$ modes, in which Ni and O sublattices move against each other along the $c$ direction and Bi not at rest. This phonon mode is polarized along the $c$ direction with the calculated displacement eigenvector of [δ(Bi) = −0.190, δ(Ni) = −0.325, δ(O) = +0.655]. It is this unstable phonon that drives a ferroelectric structural transition from the nonpolar state to the polar state. To further study the dynamic stabilities of the nonpolar state $BiNiO_2$, we have also calculated the phonon spectra in the entire Brillouin zone. The dispersion curves along high symmetry directions are plotted in Fig. 3(a). The nonpolar phase also exhibits imaginary phonon frequencies at the A(0.5, 0.5, 0.5) point. Thus, the *P*4/*mmm* $BiNiO_2$ is dynamically unstable that is the same as the nonpolar high-symmetry reference state of other typical ferroelectrics such as $BaTiO_3$.

By freezing the phonon with an imaginary frequency at the Brillouin zone center, it is possible to convert the nonpolar state into a dynamically stable polar state. We add displacements to atoms in the nonpolar state $BiNiO_2$ according to the calculated displacement eigenvector. Then the new structure is fully relaxed. The LDA calculations give $a$ = 3.787 Å, $c$ = 3.544 Å. Bi, Ni, O1, and O2 atoms site at the 1$b$ (1/2, 1/2, 0.5709), 1$a$ (0, 0, 0.0482), and 2$c$ (0, 1/2, 0.9554) (1/2, 0, 0.9554) Wyckoff positions respectively. While PBE gives optimized structural parameters of $a$ = 3.893 Å, $c$ = 3.702 Å with Bi, Ni, O1, and O2 atoms locating at the 1$b$ (1/2, 1/2, 0.5938), 1$a$ (0, 0, 0.0392), and 2$c$ (0, 1/2, 0.9484) (1/2, 0, 0.9484) Wyckoff positions respectively. The calculated structural parameters of polar $BiNiO_2$ are collected in Table I. The resulting state is a polar phase with the polar space group *P4mm*. In Fig. 2(b) we show the atomic structures of the polar $BiNiO_2$. The relative displacements along the $c$ direction between Ni and O are clear.

The polar $BiNiO_2$ with *P4mm* symmetry has 3 acoustic modes and 9 optical modes at the Brillouin zone center. Optical modes can be expressed as

$$3E + B_1 + 2A_1, \qquad (2)$$

where the three two-fold degeneracy $E$ modes and two $A_1$ modes are both infrared and Raman active, while one $B_1$ mode is Raman active. The calculated frequencies of modes at the Brillouin zone center are given in Table II. There are no modes with imaginary frequencies. The phonon spectra of polar $BiNiO_2$ are shown in Fig. 3(b). Obviously, there are no imaginary phonon frequencies through the entire Brillouin zone, which indicates the dynamic stabilities of the polar state $BiNiO_2$. The energy difference between the nonpolar and polar states is 0.13 eV and 0.19 eV for LDA and PBE calculations respectively, which are comparable to typical ferroelectrics (0.2 eV for $PbTiO_3$). This suggests the possibility of realistic switching of polarizations in $BiNiO_2$. The soft phonon mode $A_{2u}$ in the nonpolar state is energetically favorable to drive a ferroelectric structural transition from the nonpolar high-symmetry state with the *P4/mmm* symmetry into a polar *P4mm* state. We conclude that *P4mm* $BiNiO_2$ is a ferroelectric material with the *P4/mmm* state as its high symmetry paraelectric state.

This is a new type of crystal structure exhibiting ferroelectricity, which can be called the *P4mm* BiNiO$_2$ structure type.

### B. Electronic structures of nonmagnetic states

According to the previous study on strongly correlated NdNiO$_2$ and perovskite nickelates *R*NiO$_3$ (R is Bi or a rare-earth metal), the HSE hybrid functional method was essential to obtain reliable theoretical results which are consistent with experimental data [22]. Hence, we use the HSE hybrid functional to calculate electronic structures of BiNiO$_2$. In Fig. 4(a) and Fig. 5(a) we show the HSE06 band structures of nonpolar and polar BiNiO$_2$ in the nonmagnetic state. For the nonpolar BiNiO$_2$, the O 2$p$ bands distribute from −10 to −5 eV, while the Ni 3$d$ bands extend from −5 to 2.5 eV. The Ni $3d_{x^2-y^2}$ crosses the Fermi energy. Around the M point, Bi $6p_{x/y}$ states form electron pockets. There is strong mixing between Ni $3d_{x^2-y^2}$ orbital and Bi $6p_{x/y}$ orbitals along the X-M-Γ direction. At the M point, the Ni $3d_{x^2-y^2}$ state is located at just below the Bi $6p_z$ state. The Bi $6p_z$ state goes down along the Γ-Z direction and up along the Z-R direction, forming an electron pocket around Z. In NdNiO$_2$, the Ni $3d_{x^2-y^2}$ and Nd $5d_{z^2}$ states cross the Fermi energy. Compared to NdNiO$_2$, Bi $6p_{x/y}$ states change dispersions of the Ni $3d_{x^2-y^2}$ state notably [22,23]. The Ni $3d_{x^2-y^2}$, $3d_{xz/yz}$ and $3d_{xy}$ bands have weak dispersions along the Γ-Z direction, which show two-dimensional features. The Ni $3d_{z^2}$ state lies above the $3d_{xy}$ state different from the case in NdNiO$_2$. The Ni $3d_{z^2}$ state in NdNiO$_2$ is more dispersive along the Γ-Z direction than that in BiNiO$_2$. As shown in Fig. 5(a), the nonmagnetic state of polar BiNiO$_2$ is also metallic. The band structures around the Fermi energy can be compared with those in the nonpolar BiNiO$_2$. The electron pockets formed by Bi $6p_{x/y}$ and $6p_z$ states are raised in energy. The states at the Fermi energy are reduced for the structural transition from the nonpolar state to the polar state. Compared to nonpolar BiNiO$_2$, the Ni 3$d$ bands in polar BiNiO$_2$ are flatter.

To further understand electronic structures of nonpolar and polar BiNiO$_2$, we obtain on-site energies and hoppings parameters from the Wannier fitting of 14 bands around the Fermi energy corresponding to mainly Bi $p$ (3 states), Ni $d$ (5 states), and O $p$ (2 × 3 states) in the HSE06 calculations. The results are given in Table III. The amplitudes of hop between Ni $d$ and O $p$ orbitals for nonpolar BiNiO$_2$ are larger than these of polar BiNiO$_2$. The amplitudes of hop among O $2p_{x/y}$ orbitals are nearly identical for two phases. There is also significant hybridization between Ni $d$ and Bi $p$ orbitals. The Wannier functions are shown in Fig. 4(b) and Fig. 5(b) for nonpolar and polar BiNiO$_2$ respectively. They give $d$-like orbitals centered on Ni and $p$-like on the O and Bi sites. For the $3d_{x^2-y^2}$, orbitals, there are some O $2p$ contributions.

The electron density of polar BiNiO$_2$ in the (2 0 0) plane is plotted in Fig. 6(a). A highly asymmetric electron density distribution on Bi can be found, which looks like an asymmetric lobe. As shown in Fig. 6(b), the main peak of Bi 6$s$ states locate at around −14 eV. There are also lots of states distribute from −10 to −2 eV, where O 2$p$ bands also lie at. Thus, the "core" Bi 6$s$ states hybridize with O 2$p$ sates forming the asymmetric lobe. This is the typical mechanism of the lone pair driven ferroelectric structure transition [24]. Above results suggest that Bi not only drives the structural transition but also changes the electronic structures around the Fermi energy.

### C. Electronic structures of magnetic states

We now consider the ferromagnetic states. The energy difference between the nonmagnetic state and the ferromagnetic state of nonpolar BiNiO$_2$ is 0.54 eV calculated with the HSE06 functional. The local moment of Ni 3$d$ is 0.90 μ$_B$, similar to that in NdNiO$_2$ (0.94 μ$_B$). In Fig. 7 we plot band structures of the ferromagnetic state for nonpolar BiNiO$_2$. There is a large splitting between the spin minority and majority channels of the Ni $3d_{x^2-y^2}$ orbital (about 6 eV). For the spin minority channel, the Bi 6$p$ states are separated with occupied Ni 3$d$ sates forming an energy gap at each $k$ points. For polar BiNiO$_2$, the energy difference between the nonmagnetic state and the ferromagnetic state is 0.60 eV. The local moment of Ni 3$d$ is 0.93 $\mu_B$. The ferromagnetic state of polar BiNiO$_2$ is metallic as shown in Fig. 8. This is a state in which the ferroelectricity, metallicity and magnetism coexist. The splitting between the spin minority and majority channels of the Ni $3d_{x^2-y^2}$ orbital is about 5 eV. Compared to nonpolar BiNiO$_2$, states around the Fermi energy are significantly reduced as the result of the raising of Bi 6$p$ states in two spin channels and reducing of the Ni $3d_{x^2-y^2}$ state in the spin majority channel. For both the nonmagnetic state and the ferromagnetic state, the ferroelectric structural transition in BiNiO$_2$ always follows reducing of states at the Fermi energy, indicating the competition between ferroelectricity and metallicity.

For the spin majority channel of the polar BiNiO$_2$, occupied and unoccupied sates are only connected at the M point forming a QBCP just below the Fermi energy contributed from Bi $6p_{x/y}$ states, as can be found in Fig. 8(a). The site symmetry of the M point is $C_{4v}$, which has only one two-dimensional irreducible representation $E$ allowing the existence of QBCP [25,26]. The point group $C_{4v}$ consists of two generators including a fourfold rotation $c_{4z}$ ($-y$, $x$, $z$) around the $z$ axis and a mirror symmetry $\sigma_{v1}$ ($-x$, $y$, $z$). The representations of group generators for the two-dimensional irreducible representation $E$ are given by

$$D(c_{4z}) = \begin{pmatrix} 0 & -1 \\ 1 & 0 \end{pmatrix},$$
$$D(\sigma_{v1}) = \begin{pmatrix} 0 & -1 \\ -1 & 0 \end{pmatrix}. \tag{3}$$

The theory of invariant gives the following constraint on $k \cdot p$ model

$$D(R)H(k)D^\dagger(R) = H(R(k)), \tag{4}$$

where $D(R)$ is the representative matrix of operator $R$. Based on these we obtain the $k \cdot p$ effective Hamiltonian describing the QBCP

$$H(k) = \alpha_1(k_x^2 + k_y^2)\sigma_0 + \alpha_2(k_x^2 - k_y^2)\sigma_x + \alpha_3(k_x k_y)\sigma_z, \tag{5}$$

where $k_{x,y}$ are wave vectors related to the QBCP, $\sigma_{x,z}$ are the Pauli matrices denoting orbitals not spin, $\sigma_0$ is the $2 \times 2$ identity matrix, $\alpha_{1,2,3}$ are material-dependent real parameters, and the energy of the QBCP is set to zero for simplicity, which corresponds to that the Fermi energy of polar BiNiO$_2$ is turned to its QBCP. The eigenvalues are

$$E(k) = \alpha_1(k_x^2 + k_y^2) \pm \sqrt{\alpha_2^2(k_x^2 - k_y^2)^2 + \alpha_3^2 k_x^2 k_y^2}. \tag{6}$$

The quadratic band-crossing character can be understood from these eigenvalues. The QBCP in the spin majority channel of polar BiNiO$_2$ is protected by $C_4$ rotational symmetry.

Like NdNiO$_2$, the G-type AFM state displayed in Fig. 9(a) is found to be the lowest energy ground-state for polar BiNiO$_2$ [22]. The electronic structures are studied by setting the G-type AFM spin configurations on Ni using a $\sqrt{2} \times \sqrt{2} \times 2$ supercell. The HSE06 hybrid functional bands are plotted in Fig. 9(b). Due to the strong correlation in BiNiO$_2$, the Ni $3d_{x^2-y^2}$ bands split into lower and upper Hubbard bands separated by about 6 eV due to the AFM spin configuration. The lower Hubbard Ni $3d_{x^2-y^2}$ band is located higher than the Ni $3d_{z^2}$ band different from the case in NdNiO$_2$ where the lower Hubbard Ni $3d_{x^2-y^2}$ band is located below the Ni $3d_{z^2}$ band. This suggests different influences of Bi and Nd on the NiO$_2$ plane. Bi changes both the atomic and electronic structures of the NiO$_2$ plane. At the Γ point, the Bi $6p_z$ state forms the conduction band minimum and the Bi $6p_{x/y}$ states are just located at above it. HSE06 results indicate that the polar BiNiO$_2$ is a Hubbard Mott insulator. On the other hand, the computed electric polarization of polar BiNiO$_2$ is 0.49 C/m$^2$, which is comparable with typical ferroelectrics. The electric polarization in ferroelectrics can be switched by an applied electric field. Hence, as a multiferroic material, it may be possible to control the magnetic order in BiNiO$_2$ by an applied electric field experimentally, which can be studied detailly in future works.

## IV. CONCLUSIONS

In summary, we have predicted a material BiNiO$_2$ which undergoes a ferroelectric structure transition from the nonpolar *P*4/*mmm* phase to the polar *P*4*mm* phase. Both the nonmagnetic and ferromagnetic states are metallic for nonpolar and polar BiNiO$_2$. A rotational symmetry protected QBCP is found for the spin majority channel of the polar BiNiO$_2$ in the ferromagnetic state. The lowest energy ground-state for polar BiNiO$_2$ is a Hubbard Mott insulator with the G-type AFM spin configurations on Ni. The predicted material BiNiO$_2$ provides a good platform to study various coupled physical properties including ferroelectricity, metallicity, and magnetism. Through the substitution of Nd by Bi, we transform the high-temperature superconductor NdNiO$_2$ to a strongly correlated ferroelectric material BiNiO$_2$. This method may be used to design other interesting multifunctional materials based on known high-temperature superconductors.

## ACKNOWLEDGMENTS

This work was supported by the Advanced Talents Incubation Program of the Hebei University (Grants No. 521000981423, No. 521000981394, No. 521000981395, and No. 521000981390), the Natural Science Foundation of Hebei Province of China (Grants No. A2021201001 and No. A2021201008), the National Natural Science Foundation of China (Grants No. 12104124 and No. 12274111), and the high-performance computing center of Hebei University.

TABLE I. The lattice constants and atomic positions of nonpolar and polar states of $BiNiO_2$ calculated with LDA.

| Structural parameters | Nonpolar | Polar |
|---|---|---|
| $a$ | 3.835 Å | 3.787 Å |
| $c$ | 3.324 Å | 3.544 Å |
| Bi | $1d$ (1/2,1/2,1/2) | $1b$ (1/2,1/2,0.5709) |
| Ni | $1a$ (0,0,0) | $1a$ (0,0,0.0482) |
| O1 | $2f$ (0,1/2,0) | $2c$ (0,1/2,0.9554) |
| O2 | (1/2,0,0) | (1/2,0,0.9554) |

TABLE II. The frequencies of the zone-center phonon modes in nonpolar and polar states of $BiNiO_2$. The imaginary frequency is denoted as a negative value.

| Nonpolar | | Polar | |
|---|---|---|---|
| Model | Frequency (cm$^{-1}$) | Model | Frequency (cm$^{-1}$) |
| $E_u$ | 546 | $E$ | 530 |
| $E_u$ | 205 | $E$ | 366 |
| $B_{2u}$ | 121 | $B_1$ | 320 |
| $A_{2u}$ | 49 | $A_1$ | 319 |
| $E_u$ | 20 | $A_1$ | 134 |
| $A_{2u}$ | −215 | $E$ | 110 |

TABLE III. Calculated Wannier on-site energies and hoppings for nonpolar and polar states of $BiNiO_2$ in the HSE06 simulations. O1 bonds to Ni along the $y$ direction, and O2 bonds to Ni along the $x$ direction.

| Wannier on-site energies (eV) | Nonpolar | Polar |
|---|---|---|
| $d_{z^2}$ | −2.93 | −3.06 |
| $d_{xz/yz}$ | −3.43 | −3.45 |
| $d_{x^2-y^2}$ | −0.95 | −1.09 |
| $d_{xy}$ | −3.97 | −4.09 |
| $p_z$ O1 | −5.45 | −5.60 |
| $p_x$ O1 | −5.25 | −5.01 |
| $p_y$ O1 | −6.57 | −6.63 |
| $p_z$ O2 | −5.45 | −5.60 |
| $p_x$ O2 | −6.57 | −6.63 |
| $p_y$ O2 | −5.25 | −5.01 |
| $p_z$ Bi | 1.58 | 1.11 |
| $p_x$ Bi | 0.74 | 0.72 |
| $p_y$ Bi | 0.74 | 0.72 |
| Wannier hoppings (eV) | | |
| $d_{xy} - p_y$ O2 | −0.83 | −0.79 |
| $d_{xz} - p_z$ O2 | −0.91 | −0.77 |
| $d_{x^2-y^2} - p_x$ O2 | 1.74 | 1.63 |
| $d_{z^2} - p_x$ O2 | −0.59 | −0.48 |
| $p_y$ O2 $- p_x$ O1 | 0.18 | 0.14 |
| $p_x$ O2 $- p_x$ O1 | 0.34 | 0.32 |
| $p_y$ O2 $- p_y$ O1 | 0.34 | 0.32 |
| $p_x$ O2 $- p_y$ O1 | −0.69 | −0.70 |
| $p_z$ O2 $- p_z$ O1 | 0.13 | 0.09 |
| $d_{z^2} - p_z$ Bi | 0.10 | 0.17 |
| $d_{z^2} - p_x$ Bi | 0.30 | 0.27 |
| $d_{xz} - p_x$ Bi | 0.21 | 0.23 |
| $d_{xy} - p_x$ Bi | 0.13 | 0.10 |

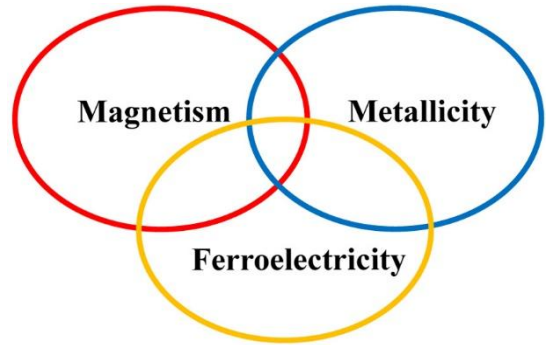

FIG. 1. The schematic plot of intertwined ferroelectricity, metallicity, and magnetism in materials.

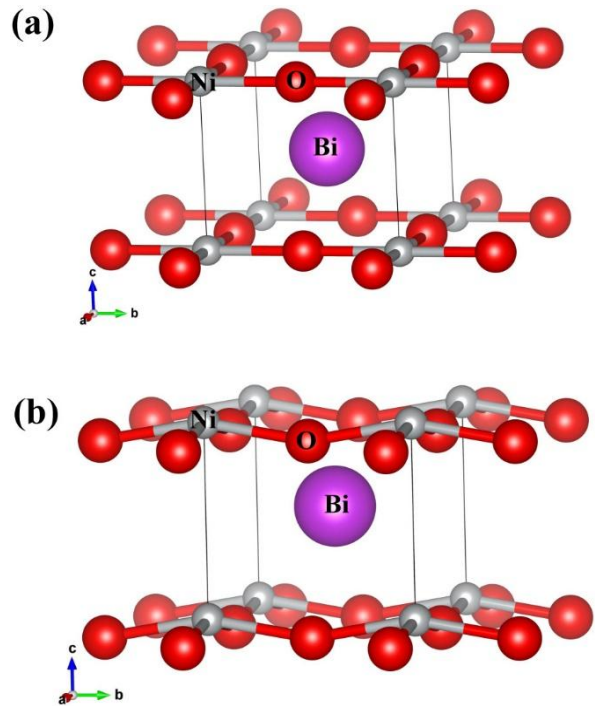

FIG. 2. The crystal structures of (a) the nonpolar $BiNiO_2$ with the $P4/mmm$ symmetry (No. 123) and (b) the polar $BiNiO_2$ with the $P4mm$ symmetry (No. 99).

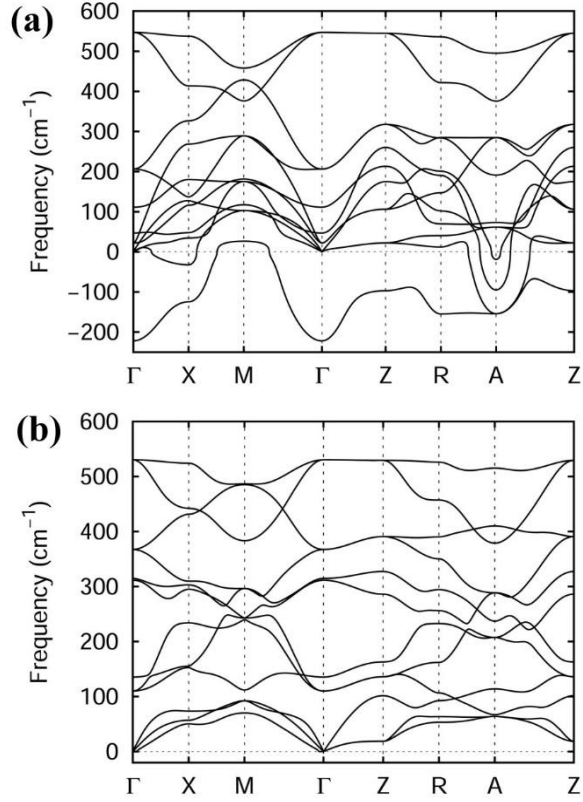

FIG. 3. Phonon spectrum of (a) the nonpolar and (b) the polar BiNiO$_2$ along the Γ(0, 0, 0)-X(0, 0.5, 0)-M(0.5, 0.5, 0)-Γ-Z(0, 0, 0.5)-R(0, 0.5, 0.5)-A(0.5, 0.5, 0.5)-Z directions.. The imaginary frequencies are indicated as negative values.

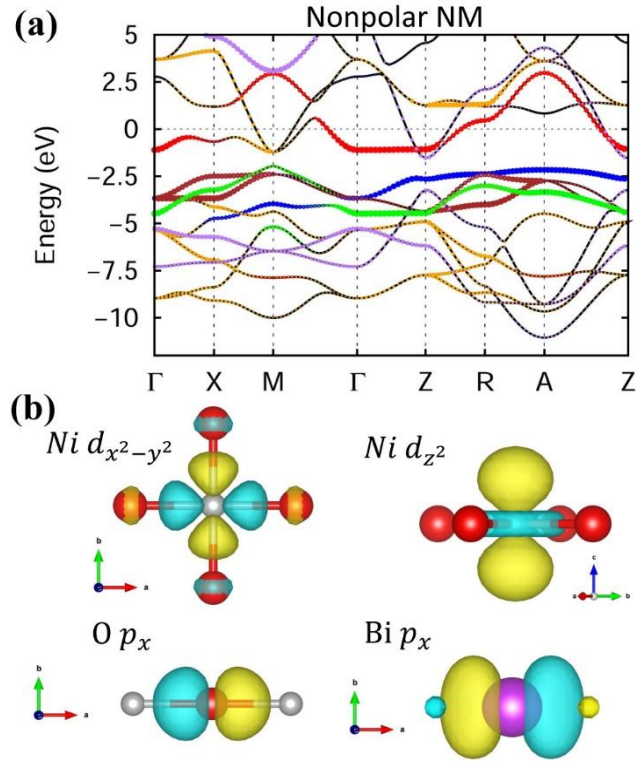

FIG. 4. (a) HSE06 band structures of the nonmagnetic state in the nonpolar BiNiO$_2$ along the Γ(0, 0, 0)-X(0, 0.5, 0)-M(0.5, 0.5, 0)-Γ-Z(0, 0, 0.5)-R(0, 0.5, 0.5)-A(0.5, 0.5, 0.5)-Z directions. The projected band structures of $d$ orbitals (red for $d_{x^2-y^2}$, blue for $d_{z^2}$, brown for $d_{xz/yz}$, and green for $d_{xy}$) in Ni and 2$p$ orbitals (orange for $p_{x/y}$ and purple for $p_z$) in Bi and O are also shown. The Fermi level is set at zero eV. (b) Wannier functions of the $d_{x^2-y^2}$, $d_{z^2}$ and $p_x$ character for nonpolar BiNiO$_2$. The sign of the Wannier function is represented by colors.

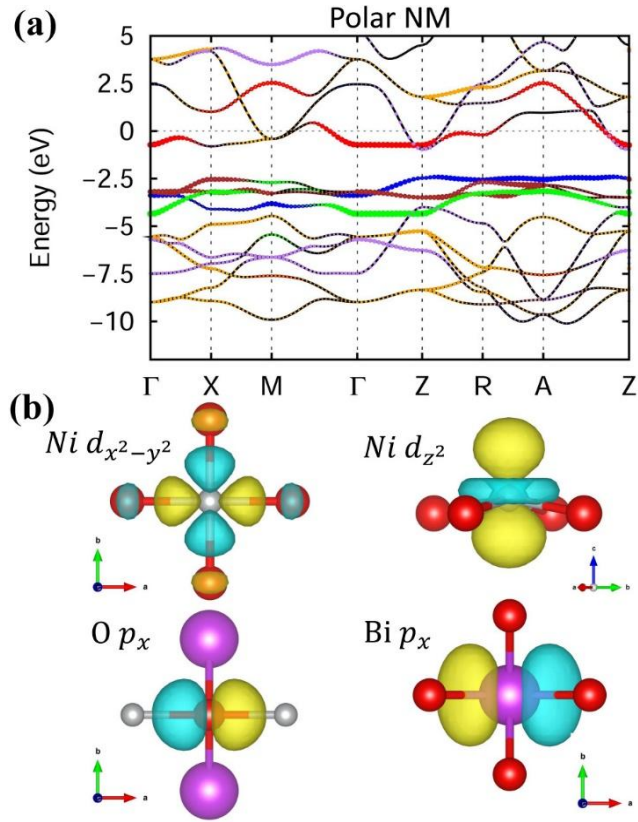

FIG. 5. (a) HSE06 band structures of the nonmagnetic state in the polar BiNiO$_2$. (b) Wannier functions of the $d_{x^2-y^2}$, $d_{z^2}$ and $p_x$ character for polar BiNiO$_2$. The sign of the Wannier function is represented by colors. The color in band structures has the same meanings as in Fig. 4.

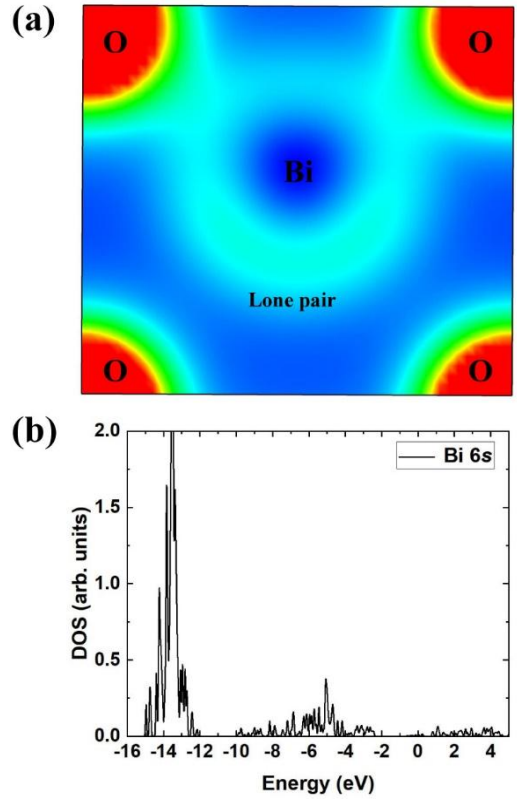

FIG. 6. (a) Electron densities and (b) partial (Bi 6$s$ orbitals) density of states for polar $BiNiO_2$.

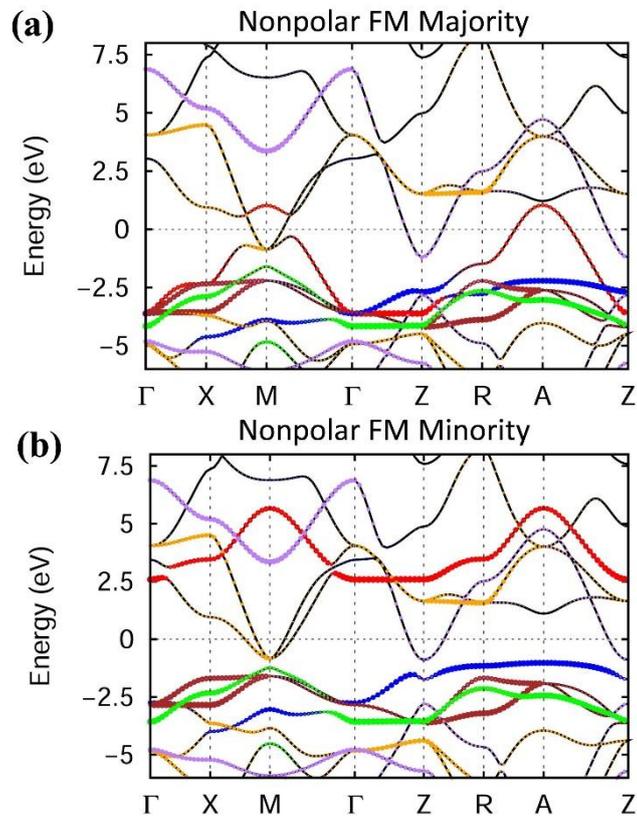

FIG. 7. HSE06 band structures of the ferromagnetic state of the spin (a) majority (b) minority

channels in the nonpolar $BiNiO_2$. The color has the same meanings as in Fig. 4.

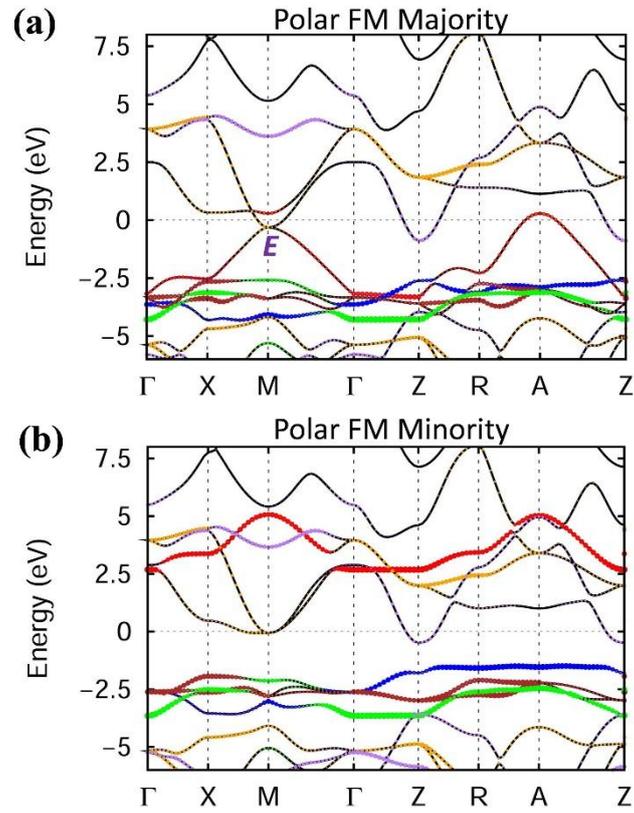

FIG. 8. HSE06 band structures of the ferromagnetic state of the spin (a) majority (b) minority channels in the polar $BiNiO_2$. The color has the same meanings as in Fig. 4.

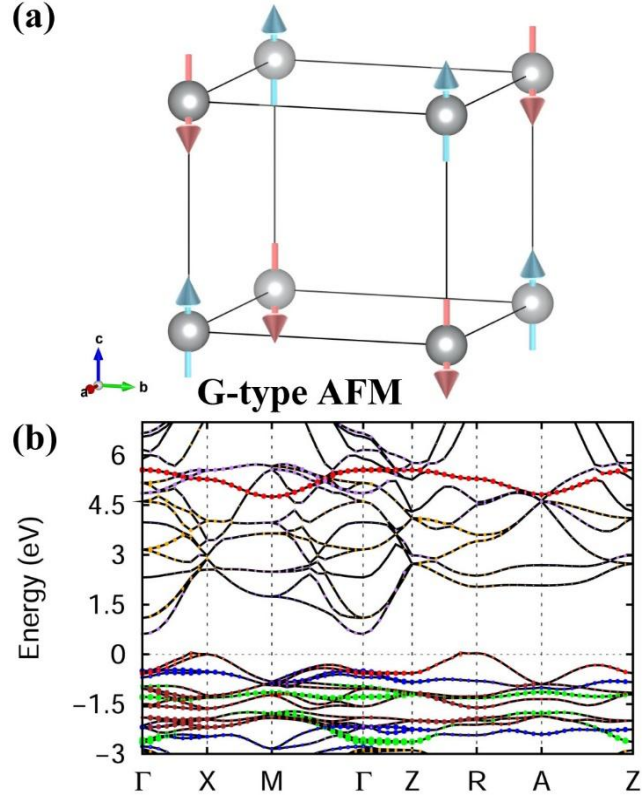

FIG. 9. (a) The G-type of antiferromagnetic (AFM) order. (b) HSE06 band structures of *G*-type AFM states for a $\sqrt{2} \times \sqrt{2} \times 2$ supercell in the polar $BiNiO_2$. The color has the same meanings as in Fig. 4.